# Potential of multi-anomalies detection using quantum machine learning

Takao Tomono
*Graduate School of Science and Technology*
Keio University
Kanagawa, JAPAN
takao.tomono@ieee.org

Kazuya Tsujimura
*Digital Innovation Divison*
TOPPAN Holdings Inc.
Tokyo, JAPAN
kazuya.tsujimura@toppan.co.jp

*Abstract*— **Maintenance of production equipment is vital in modern manufacturing. Traditional anomaly detection methods apply machine learning to vibration data, but their scalability is limited due to sensor complexity and computational constraints. In contrast, human operators often rely on auditory cues and intuition. We propose a quantum kernel-based anomaly detection method using quantum kernels in one-class SVMs to enhance feature expressiveness. Two setups were tested: (1) a miniature car track with mechanical anomalies and (2) an open-belt drive system with artificially induced anomaly sounds. Features were extracted via autoregressive (AR) model coefficients from audio. Results show quantum kernels outperform classical RBF kernels in accuracy and F1-score, particularly for varied anomalies. In one case, quantum kernels achieved sufficient classification performance, suggesting their potential for robust detection in industrial time-series data.**

## I. Introduction

In recent years, the proliferation of Internet of Things (IoT) devices has led to the accumulation of vast amounts of data across industrial sectors. Researchers and practitioners are leveraging this wealth of data to address critical challenges in manufacturing and maintenance [1], [2]. Effective maintenance of production equipment is fundamental to ensuring manufacturing efficiency and product quality. Conventional approaches rely on multiple vibration sensors attached to each machine, feeding data into machine learning models for fault detection. [3], [4], [5], [6]. As the number of devices grows, the sensor count and model training time increase dramatically, leading to high deployment and computational costs.

Traditional supervised learning models (e.g., deep neural networks) typically require thousands of labeled examples, which is infeasible in high-mix, low-volume manufacturing environments. Even when anomaly detection systems are deployed, pinpointing which machine or component is failing can be time-consuming – in practice, engineers still often walk through factories directly listening for abnormal sounds, relying heavily on personal experience and intuition.

In this paper, we propose a new approach to anomaly detection that harnesses quantum kernel methods to map time-series data into rich feature spaces. Our contributions can be summarized as follows:

**Methodology:** We present a one-class support vector machine (SVM)-based anomaly detection framework that uses quantum kernel-based feature mappings for time-series sensor data.

**Robust Feature Space:** We design a quantum feature mapping that is highly robust to noise and can capture complex patterns, enabling more effective discrimination between normal operation and multiple anomaly types.

**Multi-Anomaly Detection:** We demonstrate the potential to identify and distinguish multiple different anomaly types using a single quantum-enhanced one-class classifier, going beyond the capabilities of classical kernel methods.

Through these contributions, we aim to advance the state of the art in industrial anomaly detection and move toward realizing intelligent anomaly detection systems powered by quantum technology.

## II. Creation of Datasets

We created two datasets based on our two experimental setups. For the Open Belt Drive (OBD) system, a 5-minute audio recording of normal operation was divided into 10-second segments, yielding 30 samples of normal data. Similarly, for the Mini 4WD Mechanical (M4W) vehicle system, a 5-minute audio recording (covering multiple laps of the track) was divided into 10-second segments (each segment spans about two laps), providing 30 normal samples. Note that each lap around the M4W takes approximately 5 seconds.

Fig.1 illustrates the experimental setups for inducing anomalies. In the OBD setup (Fig. 1A), two belt drive units (one with a rubber belt and one with a metal chain belt) were equipped with wooden chopsticks as anomaly sources. Each device has pre-drilled holes, and inserting a chopstick into these rotating belts causes a sudden loud cracking sound. We defined this breaking-chopstick sound (recorded via a microphone placed near the belts) as the abnormal event for OBD. Both belt drives were triggered to produce an anomaly simultaneously for each trial.

In the M4W setup (Fig. 1B), a miniature four-wheel-drive car runs on a three-lane track. We introduced two types of anomalies on the track: wooden popsicle sticks on a section of the outer lane, and a strip of hook-and-loop fastener (Velcro) on the center lane. During a run, the car first encounters the wooden stick anomaly (producing noise when going over a step) and later the Velcro anomaly (producing a scratching sound). A

Grant [No.]: NEDO[JPNP23003], JST[JPMJPF2221]

microphone placed inside the track's loop records the running audio. These anomalies occur in sequence (outer lane then center lane) for each full circuit of the track. The continuous belt rotation noise in the OBD setup and the normal driving sound in the M4W setup constitute the normal audio for each system, while the inserted chopstick breaks and Wooden stick and Velcro contacts produce the anomaly sounds.

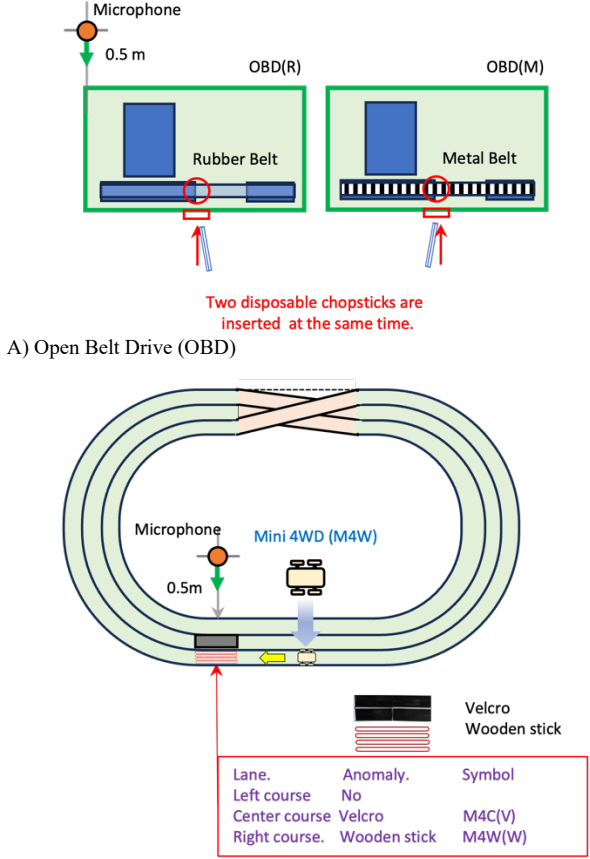

A) Open Belt Drive (OBD)

(B) Mini 4WD (M4W) Course

Fig. 1. Experimental setup for dataset creation. (A) Disposable chopsticks are simultaneously inserted into belts of two OBD systems. The belts are made by Rubber belt and metal chain. (B): M4W can drive on the course. There are two anomalies in one course: wooden popsicle sticks in the outer lanes and Velcro in the center lane. Microphone is used Directionality Microphone.

## III. QUANTUM KERNEL

Quantum kernel methods represent a promising approach to leveraging quantum computing's capabilities within the current NISQ (Noisy Intermediate-Scale Quantum) era. While classical kernels map data into higher-dimensional feature spaces to improve separability, quantum kernel approach utilize quantum state spaces that can be exponentially larger than classical counterparts.

The quantum kernel function is defined as the inner product between two quantum states:

$$\kappa(x_i, x_j) = |\langle \varphi(x_i)|\varphi(x_j)\rangle|^2 = |\langle 0|U^\dagger(x_i)U(x_j)|0\rangle|^2 \quad (1)$$

where $\phi(x_j) = |U(x_j)|0\rangle$ represents a quantum feature map that encodes classical data point $x_j$ into quantum state $|\psi(xi)\rangle$ through a parameterized quantum circuit $U(x_i)$. The feature map employs data-dependent unitary operations that create a high-dimensional representation in the quantum Hilbert space $H = (C^2)\otimes n$ for n qubits. This inner product represents the quantum state overlap and serves as a similarity measure between data points. The feature map typically employs parameterized quantum circuits with operations that create entanglement, thereby accessing feature spaces that would require exponentially many dimensions classically.

Havlíček et al.[7] introduced a framework for supervised learning using quantum-enhanced feature spaces, demonstrating that quantum kernel method could potentially offer advantages for certain classification problems. Recent theoretical work by Liu et al. [8] established rigorous conditions under which quantum kernel approach can provide provable computational advantages over classical approaches. Additionally, Huang et al. [9] explored how the power of quantum kernel scales with dataset size, showing potential advantages in the small data regime—a characteristic particularly valuable for industrial settings where anomaly data is scarce.

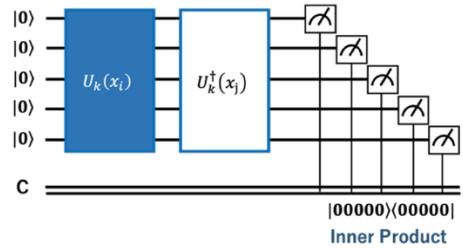

(A): Quantum circuits diagram

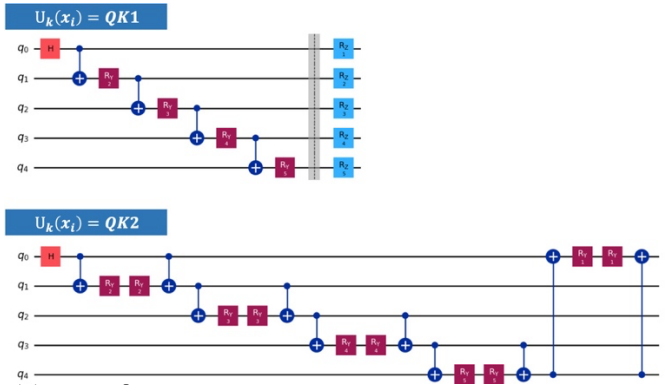

Fig. 2. Details of the quantum kernel implementation. (A) Quantum circuit diagram. (B) QK1 and QK2 architectures used in the experiment. Here, we show an example with 5 qubits.

As a practical application, it has been demonstrated that the use of quantum kernels shows high discrimination capabilities in image inspection of products in factories and in image inspection of shipments by farmers [10], [11], [12].

In this work, we focus on the expressive power of quantum kernels – their ability to construct more complex decision boundaries than classical kernels. We hypothesize that this enhanced expressive power enables a one-class SVM to detect multiple anomaly types in times series data, which is traditionally challenging for classical methods. To evaluate this, we design an anomaly detection system using quantum kernels and compare it against a classical kernel baseline. We evaluated over 25 candidate quantum kernel architectures in preliminary experiments (some from our prior work) and selected two representative kernels, dubbed QK1 and QK2, for detailed comparison. Fig.2 shows an example of the circuit-based quantum feature map we used here. We embed a 5-dimensional feature vector into a quantum state of 5 qubits (a 32-dimensional Hilbert space). Qiskit (version 0.42.0) was used for quantum simulation.

**QK1 is Linear Entanglement Kernel.** QK1 employs a linear entanglement with CNOT gates between adjacent qubits:

$$U_{QK1} = \prod_{j=1}^{n-1} CNOT_{j,j+1} \otimes_{j=1}^{n} R_y(\alpha_j x_j) \quad (2)$$

This creates a linear entanglement structure that captures correlations between neighboring features.

**QK2 is All-to-All Entanglement Kernel.** QK2 utilizes an all-to-all entanglement with CNOT gates connecting every qubit:

$$U_{QK2} = \prod_{i=1}^{n} \prod_{j=i+1}^{n} CNOT_{i,j} \otimes_{j=1}^{n} R_y(\alpha_j x_j) \quad (3)$$

This all-to-all connectivity enables simultaneous capture of higher-order correlations among all features, facilitating the identification of more complex anomaly patterns.

**QK1(Linear):** Computational complexity $\mathcal{O}(n)$, specialized for neighboring feature correlations.

**QK2(all-to-all):** Computational complexity $\mathcal{O}(n^2)$, captures global feature interactions with enhanced expressivity.

This structural difference explains QK2's superior performance in detecting subtle anomalies in the M4W dataset, where complex multi-feature correlations are essential for distinguishing between different anomaly types (wooden sticks and Velcro) from normal operation sounds.

Classical simulation complexities for n-qubit circuits are $O(2^n)$ storage per quantum state for memory and 1024-dimensional space (~4KB memory) for 10-qubits. Memory for Kernel matrix is $\mathcal{O}(N^2 2^n \cdot G)$ for N samples and G gates.

There are Scalability Trade-offs. That of QK1 is Separable structure allows $\mathcal{O}(N^2 \cdot n)$ optimization. That of QK2 is Full $\mathcal{O}(N^2 2^n)$ complexity but exponential expressivity. Quantum advantage threshold is more than 20 qubits (classical simulation >1GB). As NISQ Implementation, QK1 circuit depth is level of $\mathcal{O}(n)$ - better noise resilience. QK2 circuit depth is $\mathcal{O}(n^2)$ - requires error mitigation. Current experiments are limited to n ≲10 for tractability.

The experimental performance gains validate that quantum feature space dimensionality advantages outweigh computational overhead for our anomaly detection problem.

## IV. METHOD AND ANALYSIS

Most industrial equipment is designed to minimize failures, so labeled examples of faults are scarce. Unsupervised learning techniques are therefore commonly employed for anomaly detection. In this work, we adopt a one-class SVM approach [13], which trains only on data from normal machine operation and flags any substantial deviation as an anomaly.

**Feature extraction:** There are examples of feature extraction using AR, MFCC, and wavelets. Here, we analyze the AR model, and other examples will be reported separately. An autoregressive (AR) time series model converts the coefficients of the raw vibration signal (recording) into a feature vector, specifically, AR(p) model is represented as,

$$\mathcal{Y}_t = c + \sum_{i=1}^{p} \phi_i \mathcal{Y}_{t-i} + \epsilon_t, \epsilon_t \sim N(0, \sigma^2). \quad (4)$$

where $\mathcal{Y}_t$ is the signal at time $t$, $\phi_i$ are the model coefficients (feature values), and $\epsilon_t$ is a zero-mean white noise term accounting for unpredictable fluctuations. Here, the environmental sounds of the room and the sound of work of people were used as white noise.

We estimate the AR($p$) coefficients efficiently using the Yule-Walker equations [14] and the Levinson-Durbin recursion algorithm [15], which compute the optimal by minimizing the prediction error. Based on prior analysis, an AR order of (p=10) provided good model fidelity for our data.

**Learning pipeline:** Fig.3 outlines the overall procedure. After pre-processing and feature extraction by AR model, the dataset is split into a training set (normal data only) and a test set. The one-class SVM is then trained on the normal feature vectors. For anomaly detection, we evaluate two kernel options for the SVM.

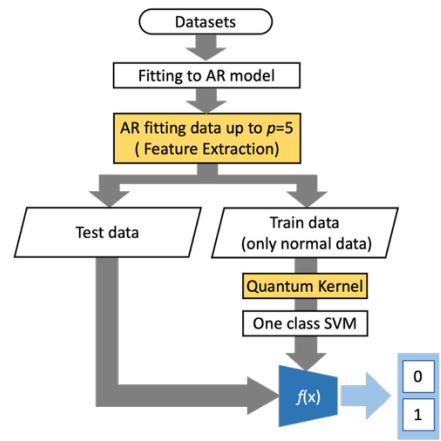

Fig.3. The flow from dataset to learning model construction and discrimination.

## V. EXPERIMENTAL RESULTS

### A. Performance vs. Number of Features

We first evaluated anomaly detection performance as a function of the feature space dimensionality (i.e. the number of AR features used). Fig.4 summarizes the one-class SVM results on the OBD and M4W datasets for different numbers of features. For the OBD dataset, with only 2 features, the classical RBF kernel outperformed the quantum kernels (achieving accuracy/F1-score ≈ 0.4, versus ≈ 0.2 for QK1 and QK2). However, as the feature count increased, the quantum kernels rapidly improved while the classical kernel lagged. QK1 and QK2 achieved sufficient classification performance (accuracy/F1-score = 1.0) using 4 and 7 features respectively, whereas the classical RBF kernel required 8 features to finally achieve an F1-score of 1.0.

On the M4W dataset, all kernels performed poorly at very low feature counts (e.g. F1-score ≈ 0.2 – 0.3 at 2-3 features). As the number of features increased, the classical RBF kernel showed almost no improvement (F1-score remain below 0.5 even at 10 features). In contrast, as the number of features on quantum kernels increased to 7, accuracy and F1-score on QK2 attained around 0.9 and these on QK1 about 0.7. These results demonstrate that the quantum kernels can extract more useful signal from the number of features than the classical kernel, especially in the more challenging M4W scenario.

The disparity between the two datasets can be attributed to the nature of their anomalies. In OBD, inserting chopsticks into the belts produces a loud, distinctive sound that is easily distinguishable from normal operation (hence even a small number of features enables high accuracy). In M4W, the anomalies (wooden sticks and Velcro on the track) generate more subtle deviations in sound, which are harder to capture; accordingly, even with more features the overall detection performance is lower.

Table 1 shows p-value on results of Fig.4. We examined the significance of QK1 and QK2 compared to RBF kernel in OBD and M4W. A statistically significant difference was observed in terms of p-value on QK1 and QK2.

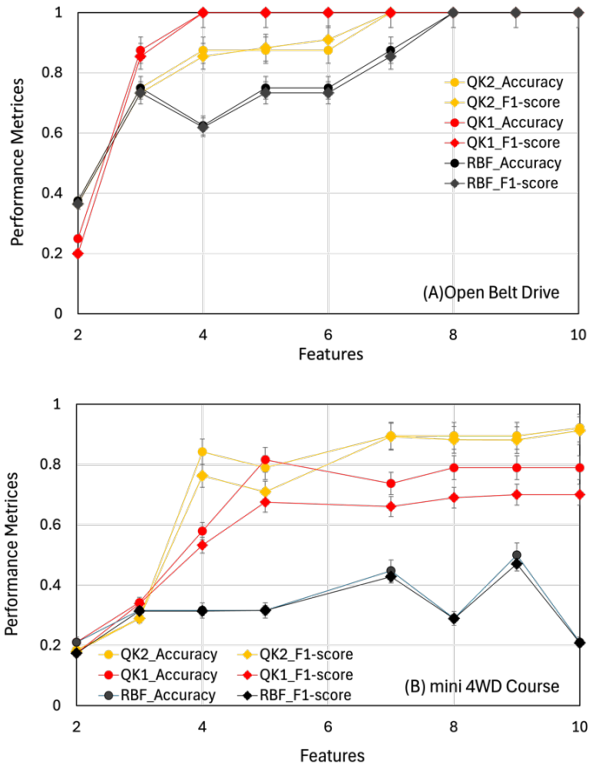

Fig.4. The relationship between performance indexes and features in OBD and M4W. Fig. (A) is OBD. Fig. (B) is M4W. The horizontal axis is the feature, and the vertical axis is the evaluation index (accuracy and F1-score).

TABLE I. P-VALUE ON RESULTS OF QK1 AND QK2 OF FIG.4 COMAPRED TO RBF KERNEL. T.TEST IS PERFORMED FROM FEATURES OF 2 TO THAT OF 10.

|  | OBD_QK1 | OBD_QK2 | M4W_QK1 | M4W_QK2 |
|---|---|---|---|---|
| P-value (vs. RBF) | 0.0420 | 0.0203 | 0.0026 | 0.0023 |

### B. Feature Space Visualization (First Two Features)

To gain insight into how the quantum kernels separate anomalies, we visualized the learned feature space in two dimensions. Fig.5. (OBD) and Fig.6. (M4W) plot the one-class SVM decision function contours for QK1 and QK2, projected onto the first and second features plane. Blue points indicate normal data and orange points indicate anomaly data.

For OBD (Fig.5), QK1 with 2 features (F2) produces a nearly linear diagonal stripe pattern in the feature space, and the normal and anomaly data points lie along the same diagonal line without clear separation. This correlates with the low F1-score (~0.2) observed at F2. By contrast, QK1 with 7 features (F7) induces a much more structured pattern: the feature space exhibits multiple separated clusters of normal data points in a wavy striped decision boundary. In this scenario, the two different anomaly types become well-isolated from the normal cluster, consistent with QK1 achieving an F1-score of 1.0 at 7 features. QK2 yields a different geometric structure: at F2, QK2's decision boundary has a horizontal striped pattern, indicating sensitivity to variations in one specific feature direction. By F7, QK2 forms a complex elliptical contour pattern with a broad decision boundary that cleanly separates normal and abnormal points.

For M4W (Fig.6.), we observed a similar trend. QK1's feature space mapping at low dimensions resulted in normal and anomaly points overlapping (leading to low initial F1-score), whereas QK2's mapping produced more complex, nonlinear decision regions that began to isolate the anomalies more effectively. At F7, QK2's contour plot for M4W showed distinct regions corresponding to the two anomaly types, whereas QK1's plot, while improved, was less clearly partitioned.

Notably, the range of the SVM decision function values differed significantly between QK1 and QK2. For QK1, the decision scores for most points were very close to zero (e.g. ±0.0002), indicating an extremely sharp decision boundary. For QK2, the score range was much wider (on the order of ±6), suggesting a more graded separation between normal and abnormal classes. A narrow score range (QK1) implies high confidence for points near the boundary but potentially less flexibility, whereas a wider range (QK2) indicates a more gradual margin that may tolerate variability at the cost of a few misclassifications (as reflected in QK2's slightly lower-than-perfect F1-score).

## VI. DISCUSSION

**Quantum Advantage and Theoretical Foundation:** The observed quantum advantage stems from the exponential dimensionality of quantum feature spaces. While classical RBF kernels operate in polynomial-scaled spaces, our quantum kernels map 5-dimensional inputs into 1024-dimensional Hilbert spaces ($2^{10}$). QK2's entangling operations create non-local correlations capturing higher-order feature interactions impossible in classical polynomial kernels. The linear separability achieved by QK1 indicates quantum rotational symmetries align with our anomaly detection geometry, while QK2's superior M4W performance demonstrates that entanglement-induced correlations are essential for distinguishing subtle anomaly patterns.

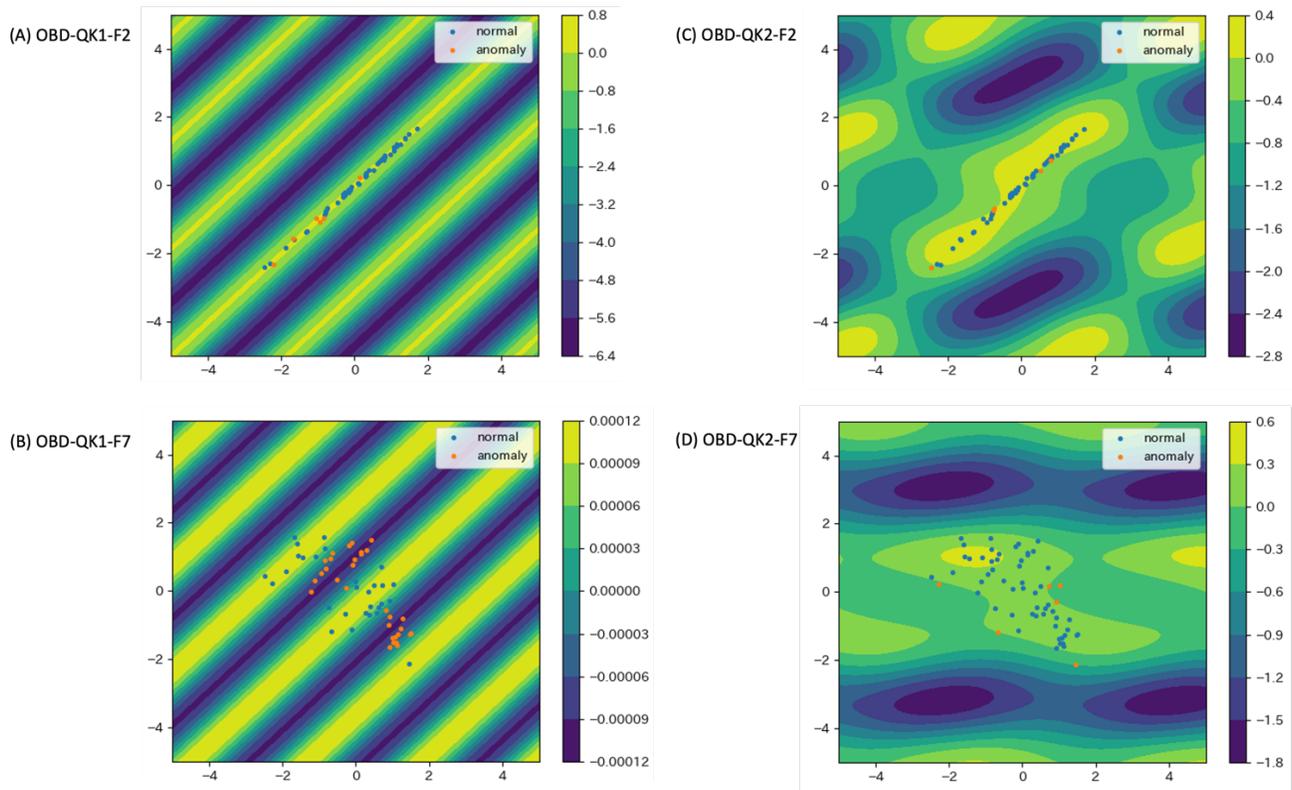

Fig.5. The mapping onto feature space of the results of one-class-SVM with quantum kernels QK1 and QK2 embedded in it for OBD

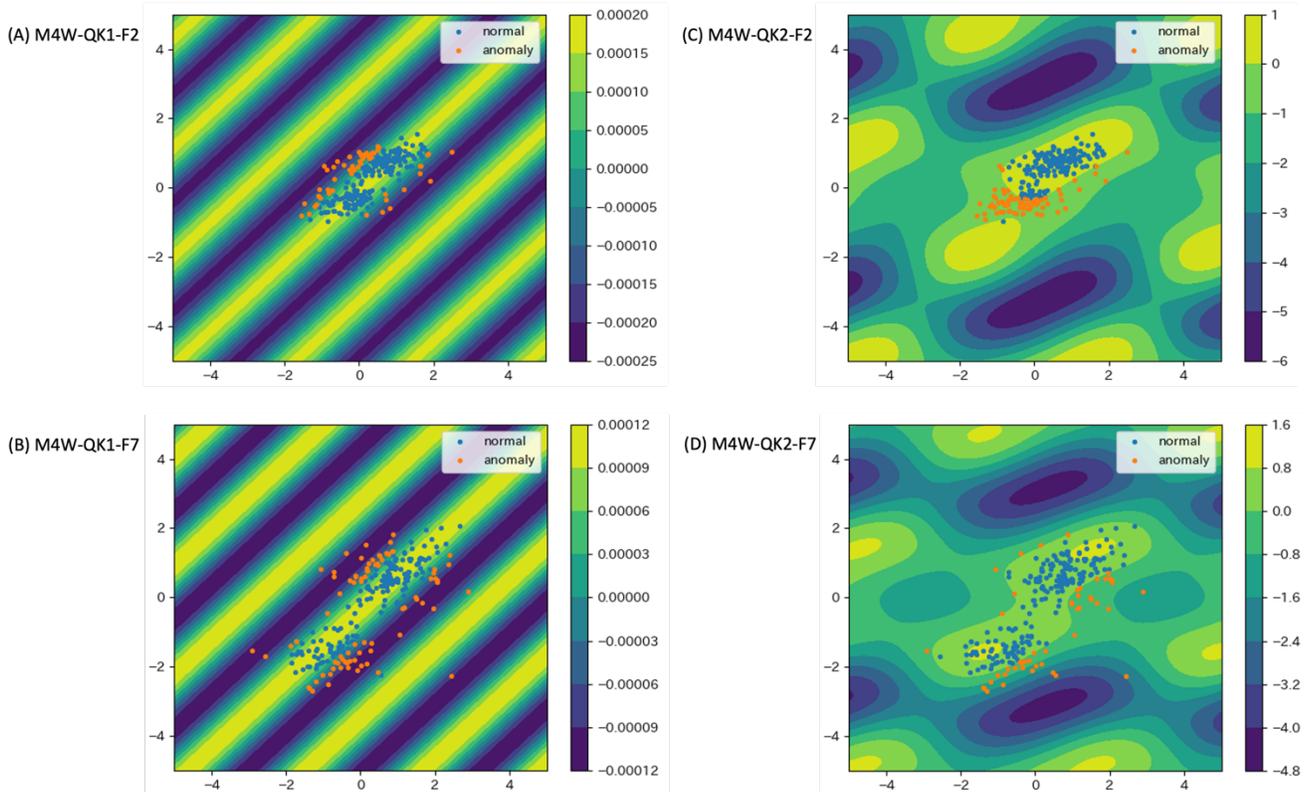

Fig.6. The mapping onto feature space of the results of one-class-SVM with quantum kernels QK1 and QK2 embedded in it for M4W.

**Comparison with Alternative Quantum ML Approaches:** Our kernel-based approach offers distinct advantages over quantum neural networks (QNNs), which require extensive training data and frequently suffer from barren plateau problems during optimization. Our one-class SVM methodology requires only normal operation data, making it ideally suited for industrial scenarios where fault examples are scarce. Unlike variational quantum algorithms that demand iterative quantum optimization processes, our method performs quantum computation exclusively during kernel matrix construction, significantly reducing overall quantum resource requirements and improving practical feasibility.

**Industrial Deployment and Practical Considerations:** Our results address critical manufacturing challenges where fault examples are inherently scarce. Achieving high accuracy with only 30 normal samples is particularly valuable for industrial settings where collecting anomaly data is costly and time-consuming. However, practical deployment faces significant challenges: current quantum circuit simulations require substantial classical resources (O ($2^{10}$) for 10-qubit QK2), potentially limiting real-time applications. NISQ device noise could degrade kernel performance, necessitating sophisticated error mitigation strategies. The 5-10 qubit range may represent the optimal balance between expressivity and noise resilience for current quantum hardware capabilities.

**Limitations and Failure Mode Analysis:** Performance variations between datasets (OBD: F1-score=1 vs. M4W:F1-score $\approx$ 0.9) indicate quantum kernel effectiveness depends heavily on underlying data structure. Our approach may struggle when anomaly types have similar acoustic signatures—visualizations show some anomaly clusters remain near normal data boundaries, indicating false negative risks. The 10-second segmentation may miss longer-term degradation patterns common in industrial equipment. Laboratory-controlled conditions may not generalize to real factory environments with varying ambient noise etc. affecting acoustic features.

## VII. Summary and Outlook

In summary, we have demonstrated a novel anomaly detection approach that uses quantum kernel methods to map time-series data into expressive feature spaces. Using two real-world setups (an OBD and a M4W), each with two different anomaly types, we showed that one-class SVMs with quantum kernels can successfully detect multiple anomalies that are difficult to distinguish with classical techniques. In particular, the quantum kernel classifiers achieved significantly higher accuracy than a classical RBF kernel for both types of anomalies in our study. The two quantum kernels studied (QK1 and QK2) exhibited distinct decision boundary characteristics, underscoring that quantum model selection must be tailored to the characteristics of the data and anomalies. Our results highlight a clear advantage of quantum kernels: they induce feature space geometries and decision boundaries that are inaccessible to classical kernels, enabling enhanced discrimination of complex anomaly patterns.

Looking forward, there are several avenues for further research. First, more advanced or hybrid quantum kernels could be developed to improve the robustness of anomaly detection in noisier real-world environments. Second, while our one-class SVM model was able to detect multiple anomaly types implicitly (by virtue of the quantum feature space causing the anomalies to form separate clusters), future work should explore methods to more explicitly distinguish each anomaly type within a unified model. Developing such techniques would move us closer to realizing quantum-enhanced smart factory.

### Acknowledgment

This work was supported by the New Energy and Industrial Technology Development Organization (NEDO) [Grant No. JPNP23003] and by the Center of Innovations for Sustainable Quantum AI (JST) [Grant No. JPMJPF2221].